# Thick ferromagnetic films and their anisotropies as described by second order perturbed Heisenberg Hamiltonian

P. Samarasekara

Department of Physics, University of Peradeniya, Peradeniya, Sri Lanka

## Abstract

Second and fourth order anisotropy dependence of energy of thick simple cubic ferromagnetic films with 10000 layers is explained using Heisenberg Hamiltonian with second order perturbation in this manuscript. The second and fourth order anisotropy constants were assumed to be constants through out the film. When the fourth order anisotropy is given by $\frac{D_m^{(4)}}{\omega}=6$, the sc(001) ferromagnetic thick films with 10000 layers can be easily oriented in 0.6 radians direction for the energy parameters given this report. Under the influence of the second order anisotropy given by $\frac{D_m^{(2)}}{\omega}=6.3$, the easy direction of sc(001) film with 10000 layers is given by 0.66 radians. Although the energy varies periodically in all cases, the maximum energy considerably decreases with fourth order anisotropy constant. According to 3-D plots, energy under influence of second order anisotropy is larger than energy under influence of fourth order anisotropy.

## 1. Introduction:

Because of the difficulties of understanding the behavior of exchange anisotropy and its applications in magnetic sensors and media technology, exchange anisotropy has been extensively investigated in recent past [2]. Due to their potential applications in magnetic memory devices and microwave devices, ferromagnetic films are thoroughly studied nowadays. Bloch spin wave theory has been previously applied to investigate the magnetic properties of ferromagnetic thin films [3]. Although the magnetization of some thin films is oriented in the plane of the film due to dipole interaction, the out of plane orientation is preferred at the surface due to the broken symmetry of uniaxial anisotropy energy. Two dimensional Heisenberg model has been used to explain the magnetic anisotropy in the presence of dipole interaction [4]. In addition to these, Ising model has been used to study magnetic properties of ferromagnetic thin films with alternating super layers [5].

The properties of ultra-thin ferromagnetic films with two and three layers have been investigated using Heisenberg Hamiltonian with second order perturbation [6]. The film with two layers behaved as an oriented magnetic film, when the anisotropy constants remained same for both layers. But the film with three layers was not equivalent to an oriented film even if the anisotropy constants remained same for all layers. The effect of few energy terms was taken into consideration for these studies. Also the energy of oriented thick ferromagnetic films was studied using Heisenberg Hamiltonian [7]. The energy found using discrete (summation) and continuous (integral) methods were exactly same. In addition to these, the energy of thick ferromagnetic films up to 10000 layers has been investigated using Heisenberg Hamiltonian with second order perturbation [8]. The matrix elements were calculated by assuming that the applied magnetic field and stress were much larger than other energy parameters. In addition, the spinel ferrite and ferromagnetic films were explained using Heisenberg Hamiltonian [9, 13, 14, 15, 16, 17]. The effect of stress induced anisotropy on coercivity was observed for rf sputtered ferrite films by us [10-12].

## 2. Model and discussion:

The Heisenberg Hamiltonian of any ferromagnetic film can be as following [6-8].

$$H = -\frac{J}{2}\sum_{m,n}\vec{S}_m.\vec{S}_n + \frac{\omega}{2}\sum_{m \neq n}(\frac{\vec{S}_m.\vec{S}_n}{r_{mn}^3} - \frac{3(\vec{S}_m.\vec{r}_{mn})(\vec{r}_{mn}.\vec{S}_n)}{r_{mn}^5}) - \sum_m D_{\lambda_m}^{(2)}(S_m^z)^2 - \sum_m D_{\lambda_m}^{(4)}(S_m^z)^4$$

$$-\sum_{m,n}[\vec{H} - (N_d\vec{S}_n/\mu_0)].\vec{S}_m - \sum_m K_s Sin2\theta_m$$

In above equation, m (or n), N, J, $Z_{|m-n|}$, $\Phi_{|m-n|}$, $\omega$, $\theta_m$ (or $\theta_n$), $D_m^{(2)}$, $D_m^{(4)}$, $H_{in}$, $H_{out}$, $N_d$, $K_s$ represent indices of layers, total number of layers, spin exchange interaction, number of nearest spin neighbors, constants arisen from partial summation of dipole interaction, strength of long range dipole interaction, azimuthal angles of spins, second order anisotropy, fourth order anisotropy, in plane applied field, out of plane applied field, demagnetization factor and the stress induced anisotropy factor, respectively.

The equation of energy can be obtained as [8],

$$E(\theta) = -\frac{J}{2}[NZ_0 + 2(N-1)Z_1] + \{N\Phi_0 + 2(N-1)\Phi_1\}(\frac{\omega}{8} + \frac{3\omega}{8}\cos 2\theta)$$

$$- N(\cos^2\theta D_m^{(2)} + \cos^4\theta D_m^{(4)} + H_{in}\sin\theta + H_{out}\cos\theta - \frac{N_d}{\mu_0} + K_s\sin 2\theta)$$

$$-\frac{[-\frac{3\omega}{4}(\Phi_0 + 2\Phi_1) + D_m^{(2)} + 2D_m^{(4)}\cos^2\theta]^2(N-2)\sin^2 2\theta}{2C_{22}}$$

$$-\frac{1}{C_{11}}[-\frac{3\omega}{4}(\Phi_0 + \Phi_1) + D_m^{(2)} + 2D_m^{(4)}\cos^2\theta]^2\sin^2 2\theta \quad \textbf{(1)}$$

Here $C_{11} = JZ_1 - \frac{\omega}{4}\Phi_1(1+3\cos 2\theta) - 2(\sin^2\theta - \cos^2\theta)D_m^{(2)}$

$+ 4\cos^2\theta(\cos^2\theta - 3\sin^2\theta)D_m^{(4)} + H_{in}\sin\theta + H_{out}\cos\theta - \frac{2N_d}{\mu_0} + 4K_s\sin 2\theta$

$C_{22} = 2JZ_1 - \frac{\omega}{2}\Phi_1(1+3\cos 2\theta) - 2(\sin^2\theta - \cos^2\theta)D_m^{(2)}$

$+ 4\cos^2\theta(\cos^2\theta - 3\sin^2\theta)D_m^{(4)} + H_{in}\sin\theta + H_{out}\cos\theta - \frac{2N_d}{\mu_0} + 4K_s\sin 2\theta$

For sc(001) lattice, $Z_0=4$, $Z_1=1$, $\Phi_0=9.0336$ and $\Phi_1=-0.3275$ [1].

First simulation will be performed for $\frac{J}{\omega} = \frac{D_m^{(2)}}{\omega} = \frac{H_{in}}{\omega} = \frac{H_{out}}{\omega} = \frac{N_d}{\mu_0\omega} = \frac{K_s}{\omega} = 10$.

3-D plot of energy versus angle and $\frac{D_m^{(4)}}{\omega}$ for a film with 10000 layers is given in figure 1. This graph indicates several energy minimums at different values of angles and fourth order anisotropy constants. For example, the energy becomes minimum at $\frac{D_m^{(4)}}{\omega} = 6$. The angle corresponding to this minimum can be observed from figure 2. The energy minimum and maximum can be observed at 0.6 and 2.8 radians, respectively. The curve indicates some overshooting at 2.8 radians.

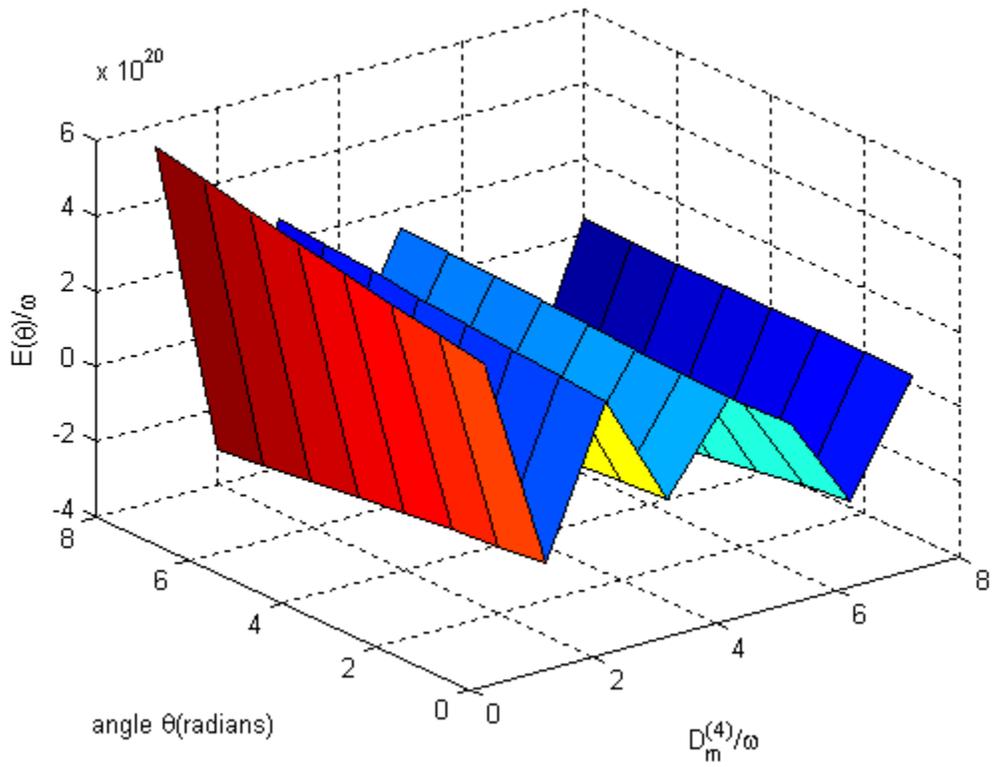

Figure 1: 3-D plot of energy versus angle and $\dfrac{D_m^{(4)}}{\omega}$ for sc(001) ferromagnetic films with 10000 layers

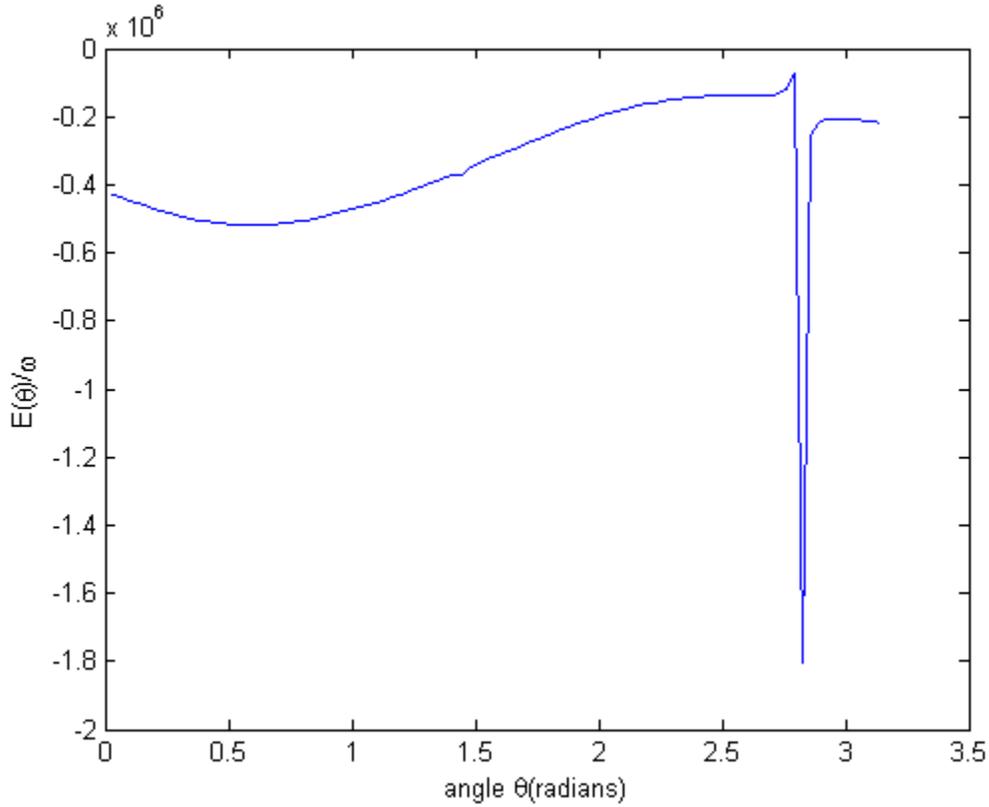

Figure 2: Energy versus angle at $\dfrac{D_m^{(4)}}{\omega}$ =6 for N=10000

When $\dfrac{J}{\omega} = \dfrac{H_{in}}{\omega} = \dfrac{H_{out}}{\omega} = \dfrac{N_d}{\mu_0 \omega} = \dfrac{K_s}{\omega} = 10$ and $\dfrac{D_m^{(4)}}{\omega}$ =5, the 3-D plot of energy versus angle and $\dfrac{D_m^{(2)}}{\omega}$ can be plotted as shown in figure 3 for a film 10000 layers. This graph also shows several energy minimums indicating that film can be easily oriented in these directions under the influence of these second order anisotropy constants. But the maximum energy gradually decreases with second order anisotropy constant. One minimum can be observed at $\dfrac{D_m^{(2)}}{\omega}$ =6.3 and, the angle corresponding to this minimum can be obtained from figure 4. Minimum and maximum energy can be observed at 0.66 and 2.85 radians, respectively.

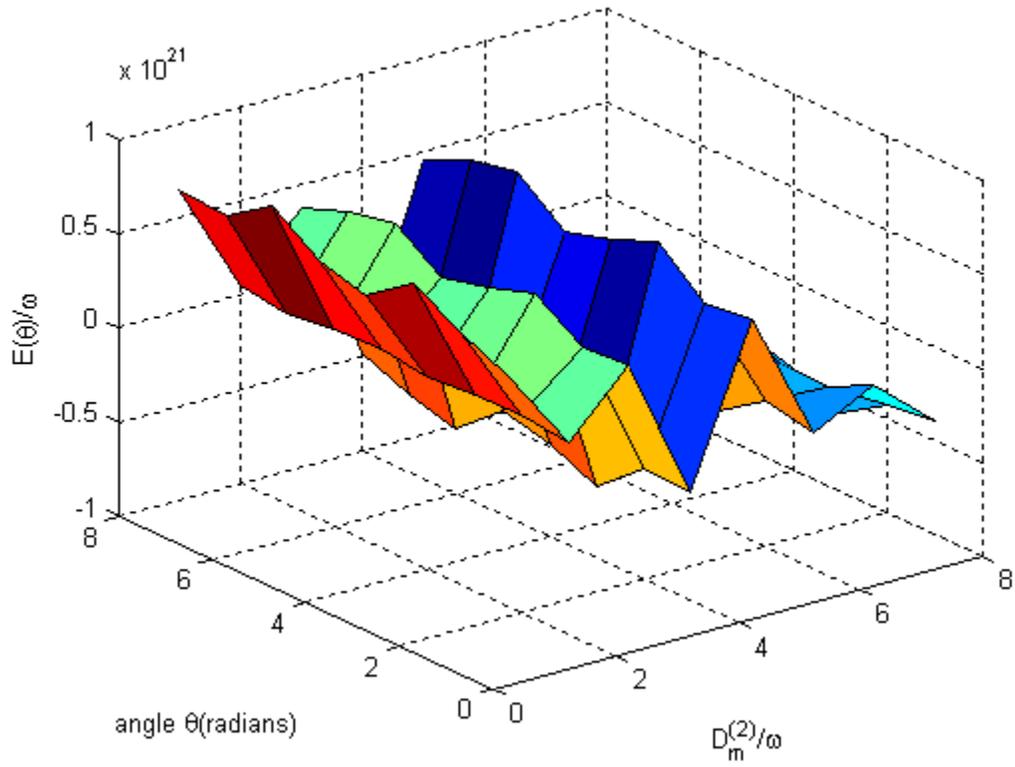

Figure 3: 3-D graph of energy versus angle and $\dfrac{D_m^{(2)}}{\omega}$ for sc(001) ferromagnetic films with 10000 layers

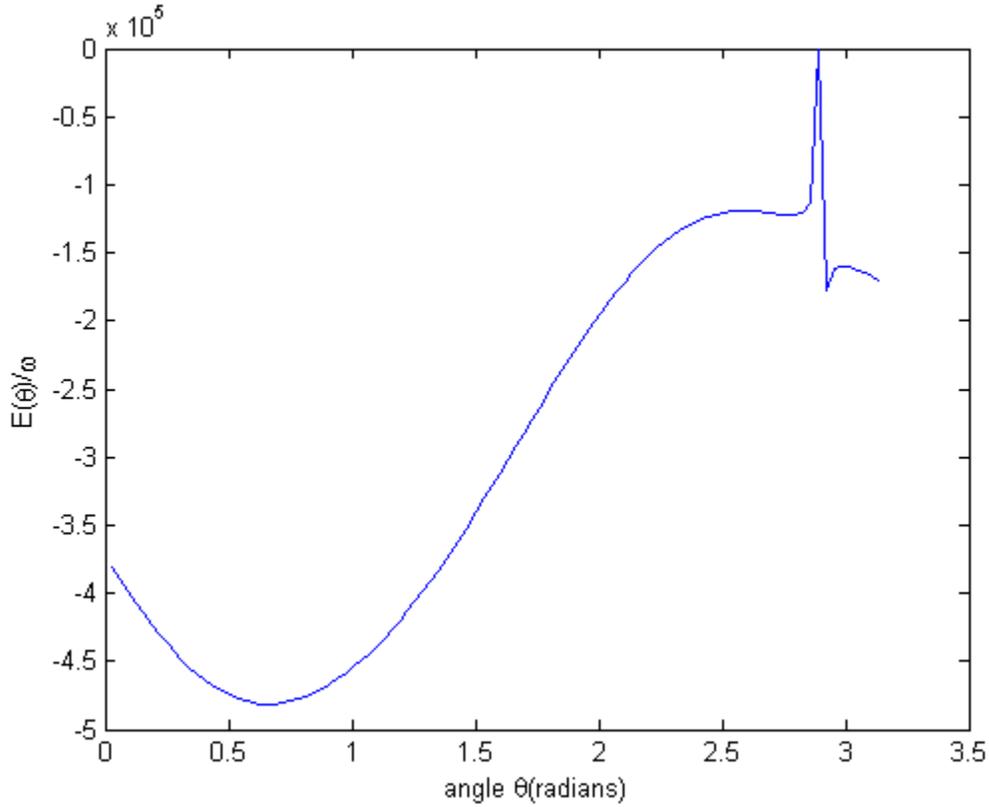

Figure 4: Energy versus angle at $\dfrac{D_m^{(2)}}{\omega}$ =6.3 for N=10000

## 3. Conclusion:

The 3-D and 2-D plot of energy curves indicate a periodic variation. This implies that the film can be easily oriented in some directions under the influence of certain values of second or fourth order anisotropies. The sc(001) ferromagnetic thick films with 10000 layers can be easily oriented in 0.6 radians direction under the influence of fourth order anisotropy given by $\dfrac{D_m^{(4)}}{\omega}$=6. When the second order anisotropy is given by $\dfrac{D_m^{(2)}}{\omega}$=6.3, the easy direction of sc(001) film with 10000 layers is given by 0.66 radians.

Although this simulation was performed for the energy parameters given in this report, this same simulation can be carried out for any values of energy parameters.